\documentclass[12pt,preprint]{emulateapj}
\usepackage{multirow}
\usepackage{hhline}
\usepackage{url}
\usepackage{amssymb}

\def\ltsima{$\; \buildrel < \over \sim \;$}
\def\gtsima{$\; \buildrel > \over \sim \;$}
\def\lsim{\lower.5ex\hbox{\ltsima}}
\def\gsim{\lower.5ex\hbox{\gtsima}}
\def\lapp{\ifmmode\stackrel{<}{_{\sim}}\else$\stackrel{<}{_{\sim}}$\fi}
\def\gapp{\ifmmode\stackrel{>}{_{\sim}}\else$\stackrel{<}{_{\sim}}$\fi}

\def\Msun{M_{\odot}}

\def\Lsun{L_{\odot}}
\def\Rsun{R_{\odot}}

\def\sax{SAX~J1748.9-2021\,}
\def\com{COM-SAX~J1748.9-2021\,}

\newdimen\minuswidth    

\setbox0=\hbox{$-$}

\minuswidth=\wd0

\catcode`@=\active

\def@{\kern\minuswidth}

\setbox0=\hbox{\rm0}

\shorttitle{}

\shortauthors{}

\begin{document}

\title{The Optical Counterpart to the Accreting Millisecond X-ray Pulsar SAX J1748.9-2021 in the Globular Cluster NGC~6440\footnote {Based on observations collected with the NASA/ESA HST (Prop. 12517, 13410), obtained at the Space Telescope Science Institute, which is operated by AURA, Inc., under NASA contract NAS5-26555.}}

\author{ M. Cadelano\altaffilmark{1,2}, C. Pallanca\altaffilmark{1}, F. R. Ferraro\altaffilmark{1}, E. Dalessandro\altaffilmark{2}, B. Lanzoni\altaffilmark{1} and~A.~Patruno\altaffilmark{3,4}}
\affil{\altaffilmark{1} Dipartimento di Fisica e Astronomia,  Universit\`a di Bologna, Via P. Gobetti 93/2, I-40129 Bologna,  Italy }
\affil{\altaffilmark{2} INAF - Osservatorio Astronomico di Bologna,  Via P. Gobetti 93/3, I-40129 Bologna, Italy }
\affil{\altaffilmark{3} Leiden Observatory, Leiden University, Neils Bohrweg 2, 2333 CA, Leiden, The Netherlands}
\affil{\altaffilmark{4} ASTRON, the Netherlands Institute for Radio Astronomy, Postbus 2, 7900 AA, Dwingeloo, The Netherlands}

\begin{abstract}
We used a combination of deep optical and $H\alpha$ images of the Galactic globular cluster NGC 6440, acquired with the Hubble Space Telescope,
to identify the optical counterpart to the accreting millisecond X-ray pulsar \sax during quiescence.
A strong $H\alpha$ emission  has been detected from a main sequence star (hereafter \com)  
located at only $0.15"$ from the nominal position of the X-ray source. The position of the star  also  agrees with the optical counterpart found by \citet{verbunt00} during an outburst.  We propose this star as the most likely optical counterpart to the binary system. By direct comparison  with  isochrones,   we estimated that \com  has a mass of $0.70 \Msun - 0.83M_{\odot}$, a radius of $0.88\pm0.02 \Rsun$ and a superficial temperature of $5250\pm80$ K. These parameters combined with the orbital characteristics of the binary suggest that the system is observed at a very low inclination angle  ($\sim8^{\circ}-14^{\circ}$) and that the star is filling or even overflowing its Roche-Lobe. This, together with the equivalent width of the $H\alpha$ emission ($\sim20$ \AA), suggest possible on-going mass transfer. The possible presence of such a on-going mass transfer during a quiescence state also suggests that the radio pulsar is not active yet and thus this system, {\ despite its similarity with the class of redback millisecond pulsars, is not a transitional millisecond pulsar}.

\end{abstract} 

\keywords{Binaries: Individual: SAX J1748.9-2021, Globular clusters: Individual: NGC 6440, Techniques: photometric}

\section{INTRODUCTION}\label{intro}
Accreting Millisecond X-ray Pulsars (AMXPs) are a sub-group of transient low-mass X-ray binaries that show, during outbursts, X-ray pulsations from a rapidly rotating neutron star. During these outbursts, the matter lost from the companion star via Roche-Lobe overflow is channeled { down} from a truncated accretion disk onto the neutron star magnetic poles, producing X-ray pulsations at frequencies $\nu \geq 100$ Hz \citep[see][and references therein]{patruno12}.  The total number of such systems currently known is 19 \citep{patruno12,sanna16b,strohmayer17}: all of them are compact or ultra-compact binaries with orbital periods usually much shorter than one day and companion stars with masses usually smaller than $1M_{\odot}$.   Two of these systems are located in the globular cluster NGC 6440, the only cluster, together with NGC 2808 and M28, known to host AMXPs. NGC 6440 is located in the Galactic bulge, above the Galactic plane, at $8.5$ kpc from the Sun \citep{valenti07}. It is a metal-rich system ($[Fe/H]\sim-0.5$, \citealt{origlia97, origlia08a}) affected by a quite large and differential extinction, with a mean color-excess $E(B-V)=1.15$ \citep{valenti04}. The cluster also hosts six (classic) radio millisecond pulsars \citep{freire08}.

\sax was discovered with the {\it Beppo}SAX/WFC in 1998 as a part of a program aimed at monitoring the X-ray activity around the Galactic center \citep{intzand99}. Since its discovery, it has experienced four outbursts, approximately one each five years: in 2001 \citep{intzand01}, 2005 \citep{markwardt05}, 2010 \citep{patruno10} and finally in 2015 \citep{bozzo15}. The X-ray pulsar has been observed pulsating at a spinning frequency of $\sim 442$ Hz and these pulsations have been used to obtain a phase-coherent timing solution \citep{gavriil07,altamirano08,patruno09,sanna16a} which revealed that \sax is a binary system with an orbital period of $\sim8.76$ hours, a projected semi-major axis of $\sim0.4$ light-seconds and a companion mass of at least $0.1 M_{\odot}$. \citet{altamirano08} suggested that the companion star is more likely a $0.85 M_{\odot}-1.1 M_{\odot}$ star, i.e. a bright main sequence or a slightly evolved star, thus implying a binary system seen at a low orbital inclination angle.  \citet{verbunt00} identified  the optical counterpart during the 1998 outburst as a blue star with $B\simeq22.7$. This outburst counterpart was identified through images obtained with a ground-based telescope and during non optimal seeing conditions. The optical emission during the outburst is dominated by the accretion processes around the neutron star, while it is expected to be dominated by the companion star during quiescence. Therefore, the identification of the optical counterpart during quiescence is needed in order to constrain the physical properties of the companion star.

The orbital properties of \sax are similar to those observed in the emerging class of ``transitional millisecond pulsars'' \citep[tMSPs;][]{archibald09,papitto13,bassa14,roy15}: binary systems that alternate stages of classical rotation-powered emission, where   radio emission is detected as in a common eclipsing millisecond pulsar (the so-called ``redback'' systems), to stages of accretion-powered emission where the radio emission is off and  X-ray pulsations are detected like in AMXPs. Optical observations of tMSPs show companion stars irradiated by the accelerated neutron star, with strong emission lines that are observed only during outbursts, marking the presence of an accretion disk around the pulsar \citep[see, for example, the cases of PSR J1824-2452I, PSR J1023+0038 and XSS J12270-4859;][]{pallanca13,patruno14,takata14,demartino15}. The similarity between \sax and the class of tMSPs suggests that this system might be a tMSP whose radio pulsar emission during quiescence has not been revealed yet. In fact, no radio pulsed emission has been detected from this object, in spite of  the radio searches devoted to this aim \citep[see][]{patruno09}. The identification of the optical counterpart during quiescence  can provide  crucial information to understand the properties and the nature of the binary system.

In the context of a long lasting program aimed at identifying optical counterparts to millisecond pulsars in Galactic globular clusters in different stages of  their formation and evolutionary path (see \citealt{ferraro01,ferraro03,ferraro15, sabbi03,pallanca10,pallanca13,pallanca14, cadelano15a,cadelano15b}), here we present the identification of the optical counterpart to \sax during quiescence. In Section 2 the observations and the data reduction procedures are described. In Section 3 we present the identification and characterization of the optical counterpart, while in Section 4 we discuss its properties. Finally, in Section 5 we summarize our results.

\section{OBSERVATIONS AND DATA REDUCTION}

This work is based on two different datasets of images obtained with the {\it Hubble Space Telescope} (HST) using
the UVIS camera of the Wide Field Camera 3 (WFC3). The first dataset (GO12517, P.I.: Ferraro) has been obtained on July 2012 and it consists of 27 images in the F606W filter with an exposure time of 392 s each and 27 images in the F814W filter with an exposure time of 348 s each. The second dataset (GO13410, P.I.: Pallanca) has been acquired during three different epochs: October 2013, May 2014 and September 2014. Each epoch consists of 5 images in the F606W filter and exposure time of 382 s, 5 images in the F814W filter and exposure time of 222 s and 10 images in the F656N filter and exposure time of 934~s. 
 
{ We used the images processed, flat-fielded and bias subtracted (``flt'' images) by the standard HST pipeline}.  The photometric analysis has been performed using standard procedures with the software {\tt DAOPHOT II} \citep{stetson87}.  First, we corrected all the images for ``Pixel-Area-Map"\footnote{For more details see the WFC3 Data Handbook.} and then we modeled the point-spread function (PSF) of each image by selecting a sample of $\sim200$ bright and isolated stars. Then we performed a source detection analysis, setting a $3\sigma$ detection limit, where $\sigma$ is the standard deviation of the measured local background. Once a list of stars was obtained, we performed a PSF-fitting of each image, by using the {\tt ALLSTAR} and {\tt ALLFRAME} packages \citep{stetson94} following the prescriptions described in \citet{dalessandro11,dalessandro14}. Only stars detected in at least half of the images in each filter have been included in the final catalog. For each star,  the magnitudes estimated in different images
have been homogenized and their weighted mean and standard deviation have been finally adopted as the star mean magnitude and its related photometric error \citep{ferraro91,ferraro92}. Finally, by using standard procedures, instrumental magnitudes have been calibrated on the VEGAMAG system using the WFC3 zeropoints publicly available at {$http://www.stsci.edu/hst/wfc3/phot _ zp _ lbn$} \citep{ryan16}.

Since the WFC3 images suffer from geometric distortions, we corrected the instrumental positions of stars by applying the equations reported in \citet{bellini11}. In order to transform the instrumental positions to the absolute coordinate system (RA, Dec), we  used the Pan-STARRS1 catalog of stars \citep{flewelling16} reported, by means of $\sim120$ in common, onto the UCAC4 astrometric standard catalog \citep{zacharias13}. Then, this catalog has been used as reference frame to astrometrize the HST dataset, by means of $~1200$ stars in common. The resulting $1\sigma$ astrometric uncertainty is $\sim 0.15\arcsec$ both in RA and Dec, thus providing a total uncertainty of about $0.21\arcsec$.

NGC 6440 is affected by a substantial differential reddening. In order to estimate the extinction variation  within the observed field of view, we adopted a method similar to that already applied  to other clusters (see \citealt{massari12}). The detailed procedure and the reddening map will be published in a forthcoming paper (Pallanca et al. 2017, in preparation). Here we just briefly describe the procedure. We first selected a sample of reference stars  with  small photometric errors and values of the sharpness parameter (``well-fitted" stars), located along the cluster evolutionary sequences in the color-magnitude diagram (CMD). These stars have been used to built a ``reference" mean ridge line. Then, for each 
star a mean ridge line has been constructed by using the fifty ``well-fitted" stars spatially located close to it. Finally, the shift $\delta E(B-V)$  needed to { register} this ridge line to the ``reference" mean ridge line is computed. The derived extinction map  shows absorption clouds with a patchy structure,  and extinction variations as large as $\delta(B-V)=1.0$ mag have been measured. The map has been used to build the differential reddening-corrected CMD used in the following analysis.

\section{RESULTS}
\label{results}

\begin{figure*}[t]
\begin{center}
\includegraphics[width=5.9cm]{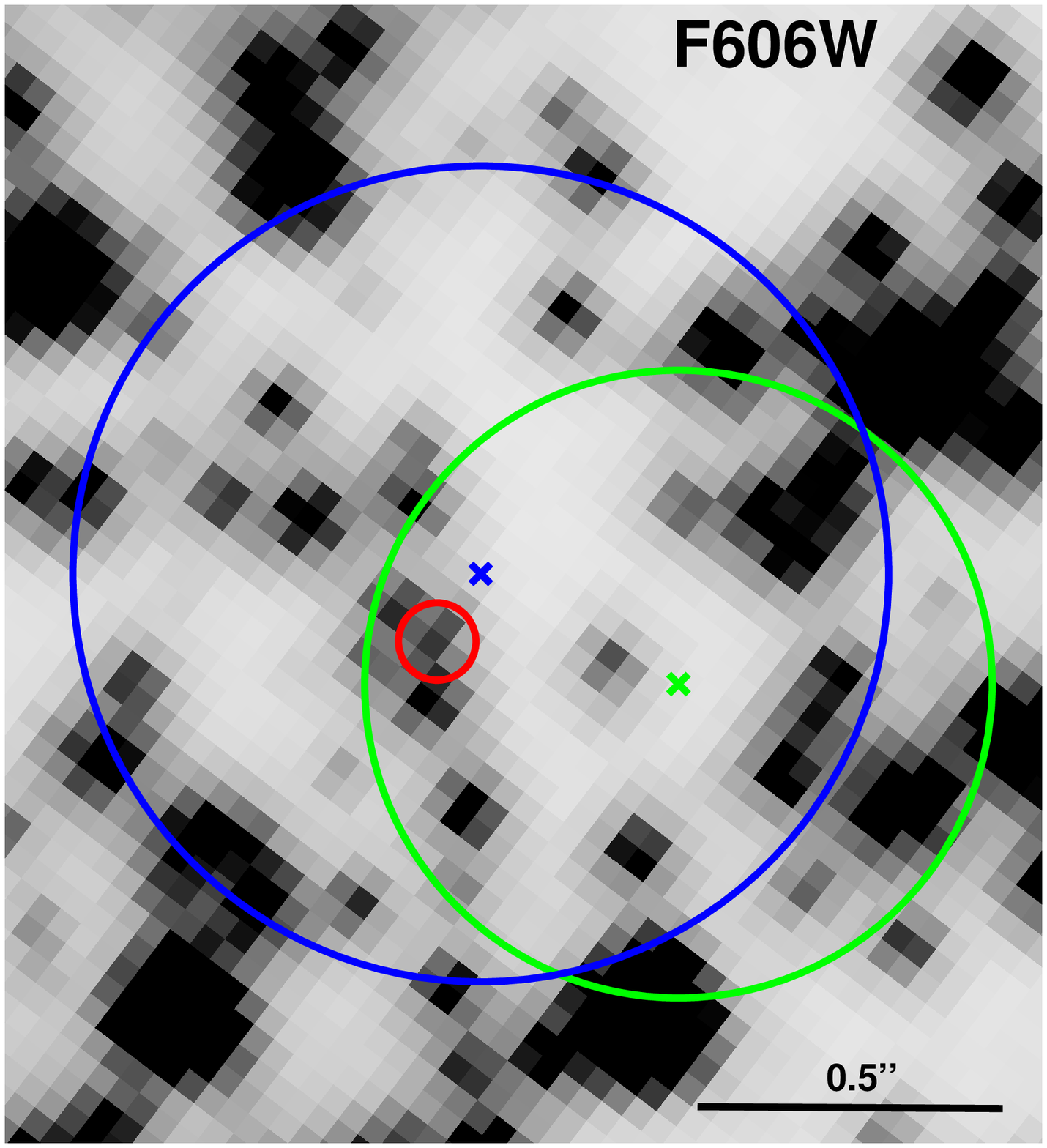}
\includegraphics[width=5.9cm]{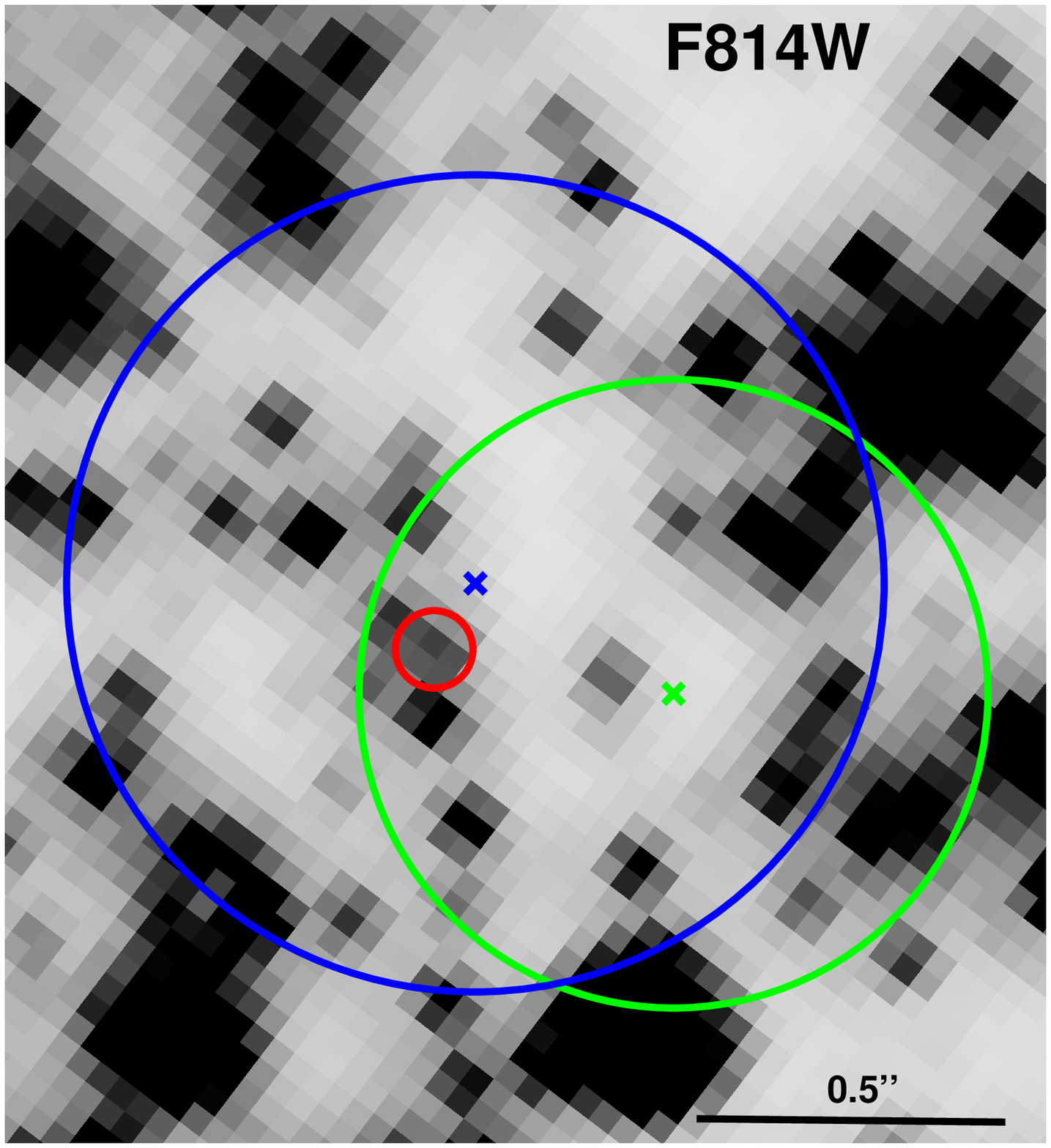}
\includegraphics[width=5.9cm]{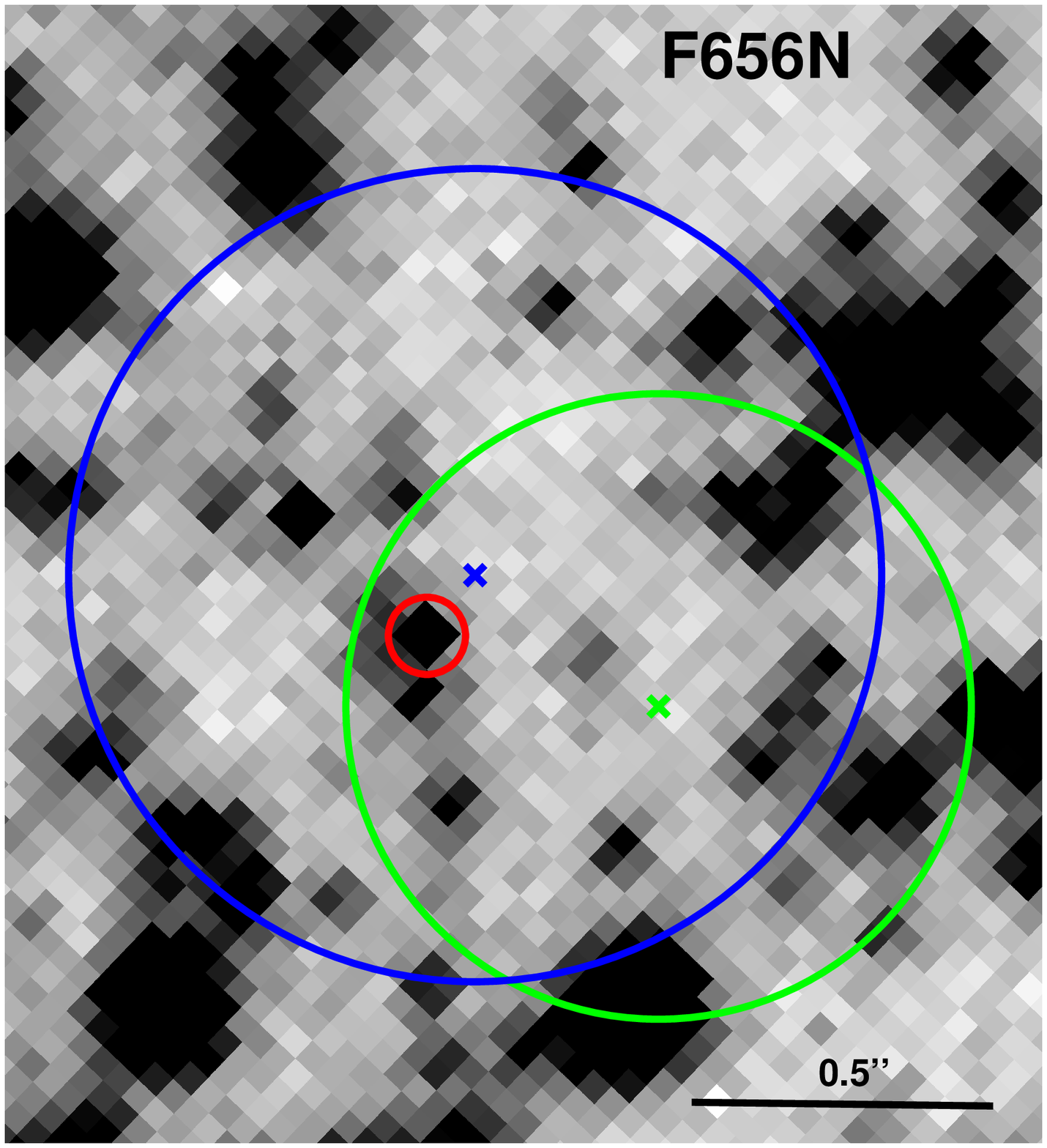}
  \caption{Finding charts of the counterpart to \com (north is up and east is left). The left, central and right panels are   combined images from the F606W, F814W and F656N expositions, respectively. In all the panels, the blue cross indicates the X-ray nominal position, while the blue circle, centered on the blue cross, has a radius equal to the combined X-ray and optical astrometric uncertainty. The green cross is centered on the outburst counterpart reported by \citet{verbunt00} and the circle has a radius equal to their astrometric uncertainty. The solid red circle marks the candidate optical counterpart. }
  \label{charts}
\end{center}
\end{figure*}

\begin{figure}[t]
\begin{center}
\includegraphics[width=8.5cm]{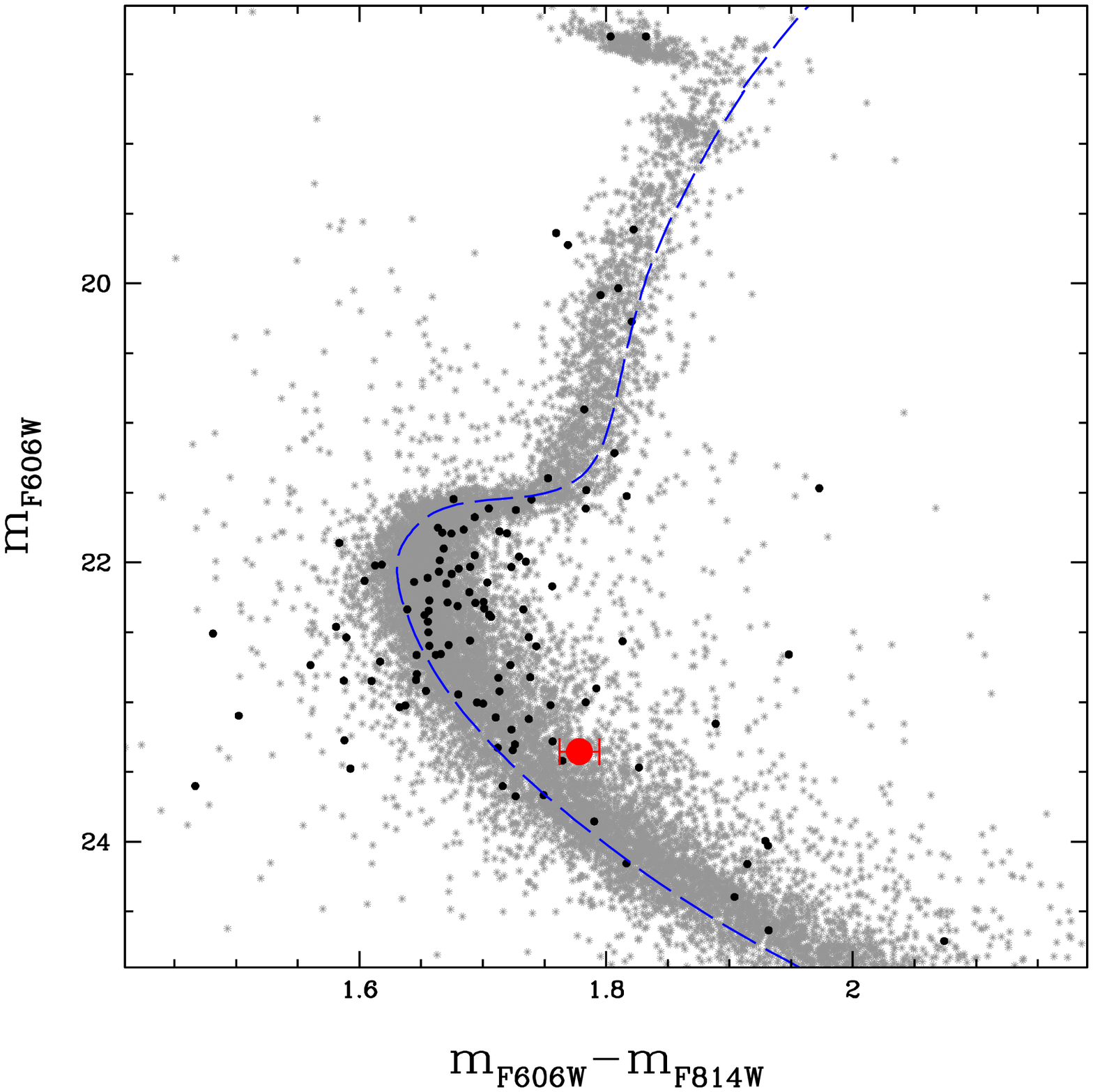}\\
\includegraphics[width=8.5cm]{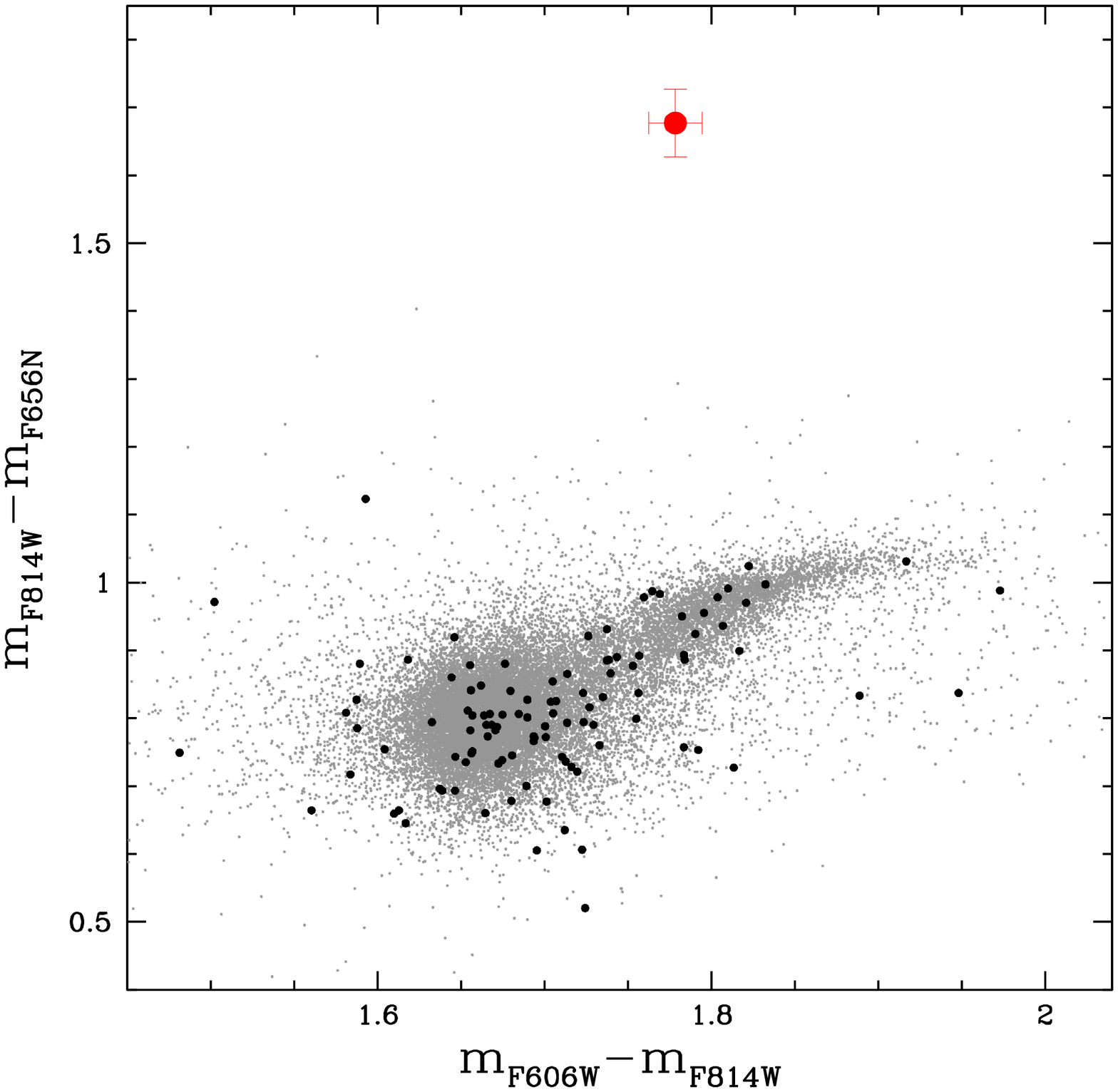}
  \caption{{\it Top:} ($\rm m_{F606W}, m_{F606W}-m_{F814W}$) differential reddening corrected CMD of NGC~6440. The stars within $2\arcsec$ from the nominal position of the X-ray source are shown as big black dots. Cluster stars detected in the WFC3 field of view are plotted in grey. The dashed blue curve is the best fit isochrone (see Section~\ref{results}) and the red point is the mean position of \com in the four epochs. {\it Bottom:} ($\rm m_{F814W}-m_{F656N}, m_{F606W}-m_{F656N}$) cluster color-color diagram. The  symbols are as in the upper panel. }
  \label{cmd}
\end{center}
\end{figure}

In order to search for the optical counterpart to \sax, we analyzed  all the objects located within a $3\sigma$ radius ($\sim2\arcsec$) from the X-ray position reported by \citet{pooley02}, where $\sigma$ is the combined X-ray ($\sim0.6\arcsec$) and optical ($\sim0.21\arcsec$) astrometric uncertainty.   The finding charts of the region around  the X-ray nominal position are shown, for all the filters,  in Figure~\ref{charts}. We emphasize that in all the epochs sampled by the observations discussed in this paper,  \sax was in a quiescence state. Thus we did not expect to find a bright star like that reported in \citet{verbunt00}. From the analysis  of the $H\alpha$ images, we found a quite promising candidate. In the bottom panel of Figure~\ref{cmd}, we show the $(m_{F814W}-m_{F656N}, m_{F606W}-m_{F814W}$) color-color diagram of the whole cluster stars (small grey dots), with all the stars detected in the region around the X-ray source position highlighted as large black dots. This diagram  has been found to be particularly powerful in pinpointing  $H\alpha$ emitters  \citep[e.g.][Pallanca et al. 2017, in preparation]{beccari14}. 
All the stars detected in the X-ray source region appear to be standard stars, with the only exception of one objects (the red dot) that shows  a quite anomalous $H\alpha$ color ($m_{F814W}-m_{F656N})\sim 1.7$,  thus indicating a  strong $H\alpha$ excess (see Section~\ref{discussion}).   This object occupies instead a standard position in the optical CMD (top panel of Figure~\ref{cmd}), being located along the cluster main sequence, about $\sim2$ magnitudes below the turn-off. 
Interestingly enough, this object is located at $\rm RA=17^{h}48^{m}52.161^{s}$ and $\rm Dec=-20^{\circ}21\arcmin32.406\arcsec$, at only $\sim0.15\arcsec$ from the nominal position of the X-ray source and it is the closest star to the X-ray position. Its location is also consistent with that of the burst counterpart proposed by \citet{verbunt00}: in fact, the distance between our and their counterpart is $\sim0.35\arcsec$, smaller than the quoted uncertainty of the latter ($\sim0.5\arcsec$). 

\begin{figure}[h]
\begin{center}
\includegraphics[width=8.5cm]{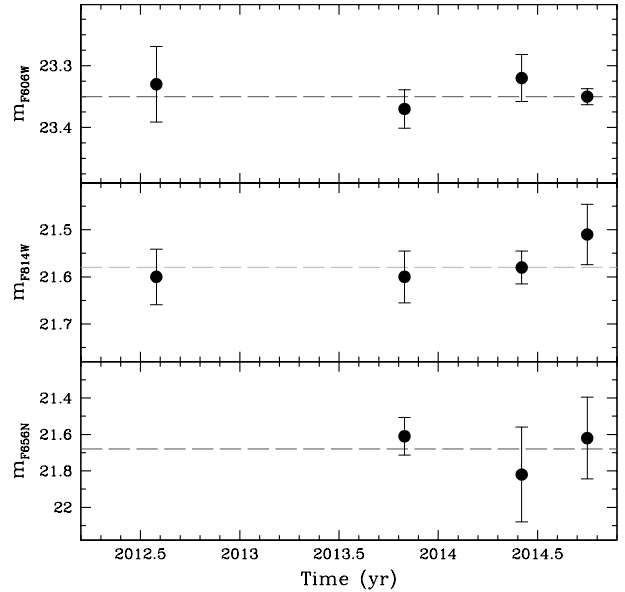}
  \caption{Magnitude of \com in the three different 
 filters measured at the epochs at which observations are available.   
The mean magnitudes are indicated with the dashed horizontal lines.  No significant luminosity variations in all the photometric bands  across the different epochs can be detected.  
}
  \label{magtime}
\end{center}
\end{figure}

Therefore,  from both the positional agreement and the presence of $H\alpha$ emission we can conclude that this object is likely the companion star to the neutron star in the binary system \sax observed during quiescence (hereafter \com).   Figure~\ref{magtime} shows the measured  magnitudes of \com in different 
 filters at the four  epochs available.   As it can be seen, no significant variation is detected across the different epochs.
The mean magnitudes in each photometric band are: $m_{F606W}=23.35\pm0.01$, $m_{F814W}=21.58\pm0.01$ and $m_{F656N}=21.68\pm0.05$\footnote{All the errors throughout the paper are quoted at $1\sigma$ uncertainty.}.  From the bottom panel of Figure~\ref{magtime} we can conclude that in the three epochs for which $H\alpha$ observations are available, a persistent  $H\alpha$ emission was present. {\ Thus, these observations  indicate an ongoing  mass transfer activity from the companion toward  the neutron star when the AMXP is in a quiescence state. This is in line with what commonly observed in different classes of interactive binaries (e.g. low-mass X-ray binaries, cataclysmic variables) that are experiencing accretion phenomena also during quiescence  \citep[e.g.][]{ferraro00,torres08,beccari14,torres14}}.

The physical properties of \com can be derived from the comparison of its position in the optical CMD with appropriate   
isochrone models. We used the  isochrone set from the Dartmouth Stellar Evolution Database \citep{dotter08}, for a 12 Gyr-old cluster \citep{origlia08b} with reddening, distance modulus and metallicity as reported in Section~\ref{intro}. The isochrone (reported as a blue dashed curve in the top panel of Figure~\ref{cmd}) nicely reproduces the cluster evolutionary sequences. By projecting the magnitude and color of \com onto the isochrone,  we found a stellar mass $M=0.73\pm0.01M_\odot$, an effective temperature of $T_e=5250\pm80 K$ and a bolometric luminosity of $0.53\pm0.01 \Lsun$, the latter two corresponding to a radius of $R=0.88\pm0.02R_\odot$ (see the discussion in Section~\ref{discussion}). From the isochrone we can also infer that the expected $B$ magnitude of the object in quiescence should be $B\simeq25.7$. This value is 3 mag fainter than that measured by \citet{verbunt00} during the 1998 outburst. Such a large variation is similar to what observed between the outburst and the quiescence states of other AMXPs \citep[see e.g.][and references therein]{patruno12} and, more generally, of transient low-mass X-ray binaries \citep[e.g.][]{ferraro15}. Since different isochrone models can lead to slightly different results, we re-made the computations by using isochrones from the $BaSTI$ database \citep{pietrinferni04} and isochrones from the Padova Stellar Evolution Database \citep{girardi00}, finding similar results.

It is worth mentioning that NGC 6440 hosts another AMXP: NGC6440X-2 \citep[][]{altamirano10}.  This is an ultracompact system with an orbital period of only $\sim0.96$ hours and a X-ray pulsar pulsating at $\sim206$ Hz. From the binary system mass function, the companion mass is expected to be $\geq 0.007 M_{\odot}$. 
Despite a careful search for the optical counterpart to this system in the available set of images, we did not find any reasonable candidate. Likely, the optical counterpart of this AMXP is still under the detection threshold, given the extremely low-mass expected for this companion star. We can therefore provide only lower-limit in luminosity for this system: $m_{F606W}>25.0$, $m_{F814W}>23.5$ and $m_{F656N}>23.0$.

\section{DISCUSSION}
\label{discussion}

\begin{figure}[t]
\begin{center}
\includegraphics[width=8.5cm]{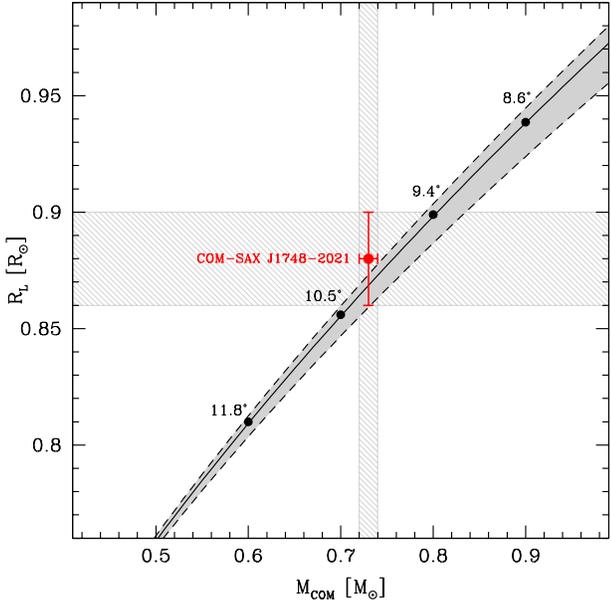}
  \caption{Roche Lobe radius as function of the \com mass. The solid line represents the analytic prediction for a pulsar mass of $1.4M_{\odot}$, while the shaded gray area surrounded by the dashed curves correspond to the predictions for a pulsar mass ranging from $1.2 M_{\odot}$ to  $2.4 M_{\odot}$. The labelled black dots indicate the inclination angle of the binary system as predicted by its mass function.   The red circle and gray striped area mark the derived \com radius and mass.}
  \label{raggio}
\end{center}
\end{figure}

{\ We can now compare the radius of \com  obtained in
Section~\ref{results}  with the expected dimension of the Roche-Lobe radius}.  The latter quantity can be estimated according to the following relation  \citep{eggleton83}:
$$
R_{L}=\frac{0.24M_{PSR}^{1/3}q^{2/3}\left(1+q\right)^{1/3}P_{ORB, hr}^{2/3}}{0.6q^{2/3}+\log(1+q^{1/3})}
$$
where $q$ is the ratio between the companion and the pulsar mass and $P_{ORB, hr}$ is the orbital period in hours. 
In Figure~\ref{raggio} (solid line) we plot the Roche-Lobe radius, computed by
assuming a pulsar mass in the range $1.2M_\odot-2.4M_\odot$, as a function of the companion mass. {\ The position of \com  in this diagram (large filled dot) indicates that it is at least completely filling its Roche-Lobe and possibly even overflowing it}. Indeed, the  Roche-Lobe radius corresponding to the  estimated mass of \com   is $0.86R_{\odot}-0.87R_{\odot}$, implying a filling factor of $0.99-1.05$. {The derived filling factor is in agreement with what expected from such a system, where the presence of on-going mass transfer implies that the companion star is most likely in a Roche-Lobe overflow state. However, it is worth noticing that the mass derived for \com from standard stellar isochrones can be biased. In fact, the companion star could have suffered a strong mass loss if it is the same object that has recycled, via mass transfer, the neutron star. Such an effect is not accounted for by the stellar evolutionary models used to create isochrones, thus introducing a bias in the derivation of the companion physical properties \citep[see the notorious cases of PSR~J1740$-$5340A and PSR~J1824$-$2452H:][]{ferraro01,pallanca10,mucciarelli13}. On the other hand, we can assume the derived photospheric radius more reliable, since it exclusively depends on the companion luminosity and temperature. Setting this measured radius equal to the Roche-Lobe radius, we found that the companion star is filling its Roche-Lobe in the mass range of $0.70\Msun-0.83\Msun$. However, the possible presence of heating of the companion star due to the neutron star emitted flux could affect the observed luminosity and temperature of the companion star, introducing an additional bias, which is difficult to quantify. Nevertheless, the position of \com, compatible with the cluster main sequence, suggests that this effect might not be very relevant for this system, at odds with what observed for strongly heated companion stars \citep[see, e.g.,][]{edmonds02,pallanca14,cadelano15a}.} We can therefore conclude that the observed properties of  \com are likely compatible with that of a binary system where the secondary star has a mass of $0.70 \Msun -0.83 \Msun$ and it is filling and possibly overflowing its Roche-Lobe, whose radius is $0.88\pm0.02R_{\odot}$. These quantities can be used to constrain the inclination angle of the system. Using the orbital solution reported by \citet{patruno09} and \citet{sanna16a}, we found that, for a pulsar mass in the range $1.2M_\odot-2.4 M_\odot$, the binary inclination angle should be very low, between $8^{\circ}$ and $14^{\circ}$. Interestingly, a  low orbital inclination angle was independently suggested by the absence of dips and eclipses in the X-ray light curve \citep[e.g.][]{sanna16a} and by the expected properties of the companion star discussed by \citet{altamirano08}.

The evidence of  $H\alpha$ emission previously discussed and shown in Figure~\ref{cmd} can be used to estimate the equivalent width (EW) of the emission line of main sequence stars. In doing this, we followed the method reported by \citet{demarchi10} and already used in previous papers \citep[][Pallanca et al. 2017, in preparation]{pallanca13, beccari14}. Briefly, the  excess in the  de-reddened $H\alpha$ $(m_{F606W}-m_{F656N})_0$ can be expressed in terms of the equivalent width of the $H\alpha$ emission by using equation (4) in \citet{demarchi10}: $EW=RW\times[1-10^{(-0.4\times \Delta H\alpha)}]$, where 

\begin{itemize}
\item $RW$  is the ``rectangular width" of the adopted $H\alpha$ filter, its definition being similar to that of equivalent width used to measure the intensity of an emission/absorption line. According to Table 4 in \citet{demarchi10}, $RW=17.48 \rm \AA$ for the HST-WFPC3 $H\alpha$ filter we adopted here.  

\item $\Delta H\alpha$ is the difference in the de-reddened $H\alpha$ color $(m_{F606W}-m_{F656N})_0$ between \com and the value expected from a star with the same optical color $(m_{F606W}-m_{F814W})_0$ but showing  
no $H\alpha$ emission. 
\end{itemize} 

\begin{figure}[b]
\begin{center}
\includegraphics[width=8.5cm]{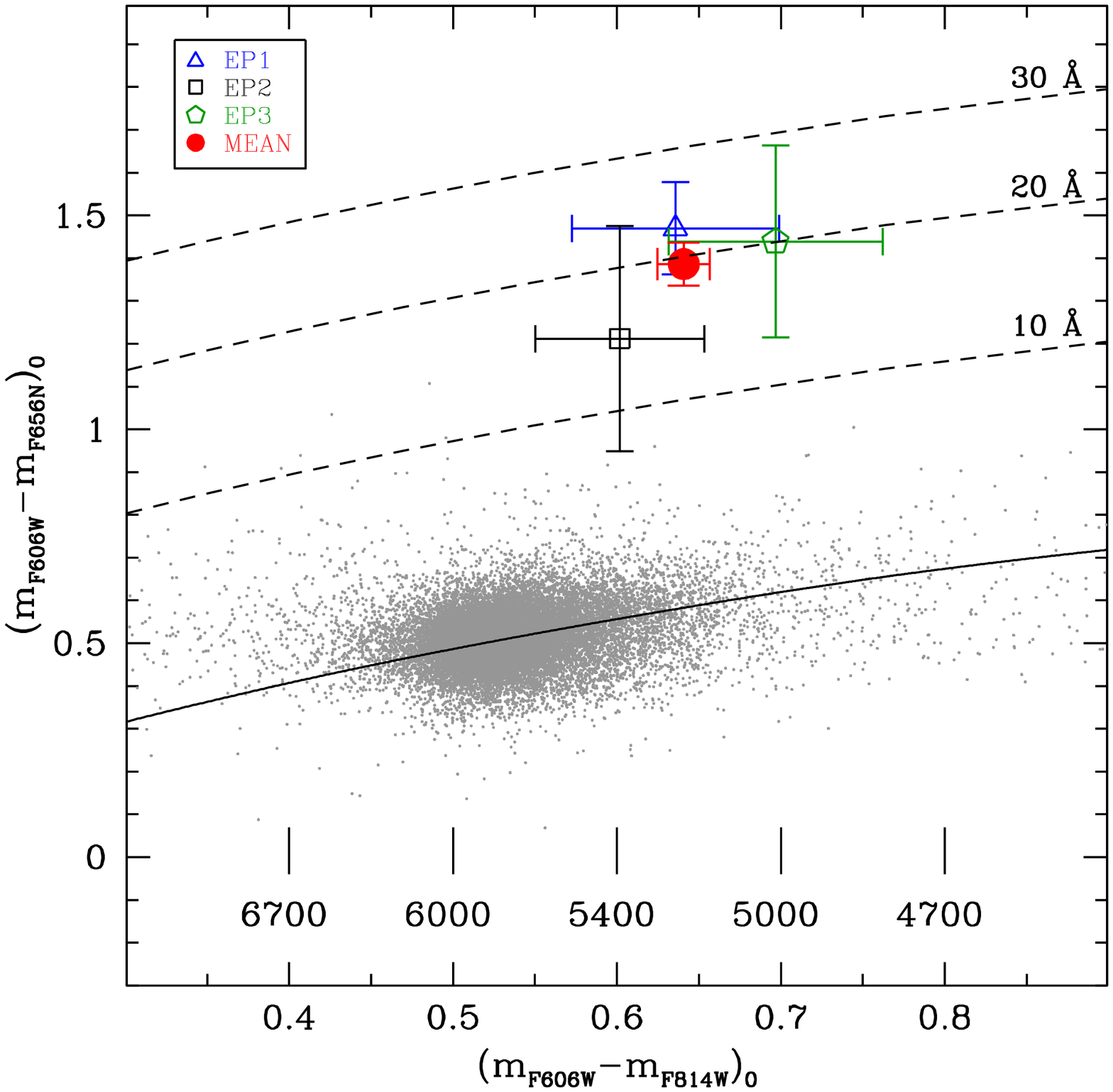}
  \caption{Reddening corrected $\rm (m_{F606W}-m_{F814W})_0$ vs $\rm (m_{F606W}-m_{F656N})_0$ color-color diagram of NGC 6440. The solid line marks the region occupied by main sequence stars with no $H\alpha$ emission, while the dashed ones show, respectively, the regions where stars with $H\alpha$ emission and an EW of 10 \AA, 20 \AA and 30 \AA are located. The colored points are the positions of \com in the different epochs, as reported in the legend. Effective temperatures (in Kelvin) related to the corresponding colors are also marked.}
  \label{colcol}
\end{center}
\end{figure}

On the basis of this relation, different curves at increasing $H\alpha$ EW can be computed and they are plotted in Figure~\ref{colcol}. When main sequence stars are plotted in this diagram, the vast majority of them are located around the ``no $H\alpha$ emission'' curve (solid curve), as expected by canonical cluster stars. \com   is instead located significantly above this line, showing a mean systematic excess of  $\Delta H\alpha = 0.80\pm0.07$ in all the sampled epochs.
As it can be seen from  Figure~\ref{colcol} such an excess corresponds to an EW of the $H\alpha$ emission  of $19\pm1$ \AA.
Such a value is too large to be attributed to chromospheric activity \citep{beccari14}. It is instead
 a typical value for system with a low mass accretion rate. In fact it turns out to be in agreement with the $H\alpha$ EWs measured in the majority of quiescent low mass X-ray binaries with a neutron star accretor \citep[$\rm EW = 20\rm \AA$-$\rm50\AA$, see][and references therein]{heinke14}. The evidence of such a low-level accretion rate in this system was already suggested by the X-ray studies of \citet{bahramian15}, performed two years after and one year before a burst. The value of $\Delta H\alpha$ just measured can be used to directly  estimate the $H\alpha$ luminosity due to the accretion processes $L(H\alpha)$, by using the photometric zeropoints and the values of the inverse sensitivity (PHOTFLAM parameter)  publicly available for all the WFC3 filters (see {$http://www.stsci.edu/hst/wfc3/phot_zp_lbn$}). At the cluster distance, we found $L(H\alpha)=1.27\pm0.08\times10^{-4} L_{\odot} $. This value, together with the \com mass and radius quoted in Section~\ref{results}, can be inserted in equation (7) of \citet{demarchi10} to estimate the mean mass transfer rate of the binary system, that turns out to be $\dot{m}\sim3 \times 10^{-10} \, \, M_{\odot} \, yr^{-1}$. 
Note, however, that the derived value of $\dot{m}$ must be taken with extreme caution, since the \citet{demarchi10}  method has been calibrated for accretion processes in pre main sequence stars, hence its applicability to the different cases (as low-mass X-ray binaries) could be risky. 

{\ \citet{sanna16a} measured for this binary system a large orbital period derivative of $1.1\times10^{-10}$. This large value is interpreted as the result of a non conservative mass transfer driven by the emission of gravitational waves. In this model, the large orbital period derivative implies a large time-averaged mass transfer rate ($\sim10^{-8} \Msun \, yr^{-1}$) that can be explained by a companion star with a low-mass of $\sim0.12 \Msun$, where only the $3\%$ of its lost mass is accreted by the neutron star (see their Figure 9). Our findings are not in agreement with such a scenario, since the companion mass is significantly more massive. If we assume that the large orbital period derivative is indeed the result of a strong mass transfer, our and their results could be reconciled by assuming that the fraction of lost mass that is accreted by the neutron star is even smaller than $3\%$. However, in the previous paragraph we roughly estimated that the mass transfer rate during quiescence is $\dot{m}\sim3 \times 10^{-10} \, \, M_{\odot} \, yr^{-1}$. This value, although extremely uncertain, is $\sim100$ times smaller than their predicted value. This could suggest that the mass transfer rate, estimated by \citet{sanna16a} on the basis of the orbital period derivative, is overestimated. More generally, the disagreement between our and \citet{sanna16a} results could suggests that the large orbital period derivative is due to a different effect such as, for example, a variable quadrupole moment of the companion star \citep{applegate92,applegate94,hartman08,patruno12b}, a phenomenon commonly invoked to explain the time evolution of the orbits of black-widows, redbacks and tMSPs \citep{applegate94,archibald13,pallanca14,pletsch15}. The similarity of SAX J1748.9-2021 to the redback class might corroborate this hypothesis, although other alternatives exist \citep[see for example][for a discussion]{patruno16}.}

\begin{figure}[t]
\begin{center}
\includegraphics[width=8.5cm]{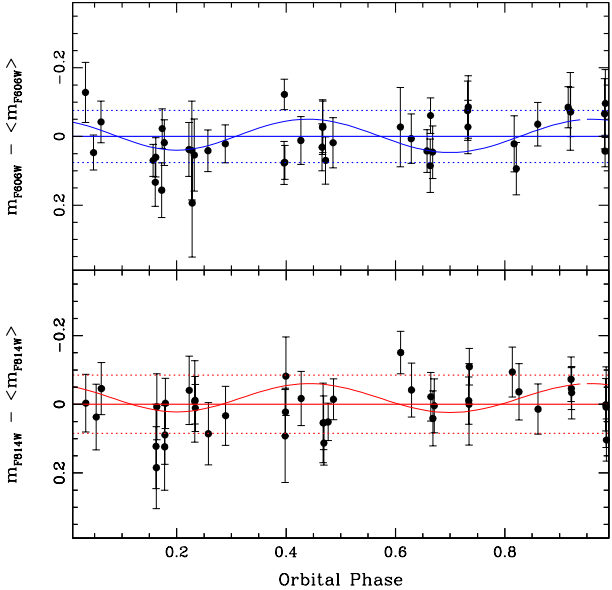}
  \caption{{ Light curve of \com (black circles and error bars) obtained by folding the F606W measurements (top panel) and F814W measurements (bottom panel) with the binary system orbital parameters. The solid and dashed horizontal lines are respectively the mean and the standard deviations of the measurements. No evidence of variability associated to the orbital period is visible with the photometric errors on the single exposures. For illustrative purposes, we also show the simulated light curves obtained with {\tt NIGHTFALL} (blue and red curves) for an orbital inclination of $30^{\circ}$, which is, however, excluded by our analysis.}}
  \label{curva}
\end{center}
\end{figure}

Since the analyzed dataset samples almost homogeneously the entire orbital period of the system,
an additional  aspect that we can investigate is the possible presence of light modulations.
Indeed, sinusoidal variations  due to irradiation processes or ellipsoidal deformation of the star, are expected to be observed in these systems \citep[see e.g.][]{homer01,davanzo09}, although the amplitude of the modulation strongly depends on the system inclination angle. In order to determine  the amplitude of the light curve expected from the system, we 
constructed a very basic model of \sax by using the software {\tt NIGHTFALL}\footnote{This software is publicly available at \url{http://www.hs.uni-hamburg.de/ DE/Ins/Per/Wichmann/Nightfall.html}.}. We simulated a set of light curve models (in the F606W and F814W filters)\footnote{Note that since the software does not allow to evaluate light curves for the specific  WFC3 photometric filters,  we  used the  V and I Johnson filter to simulate respectively the F606W and F814W light curves.} with a point-like primary star of $1.4M_{\odot}$ and a Roche-Lobe filling companion star with masses in the range of $0.1 \Msun - 1 M_{\odot}$ (compatible with both the binary mass function and the cluster stellar population)\footnote{The binary system mass function predicts, for a system with very low inclination angles ($i<5^\circ$), companion masses larger than $2\Msun$, incompatible with the old population of stars in globular clusters.}. { We found that amplitudes $\lesssim0.01$ mags are expected for an inclination angle of $\sim10^{\circ}$, corresponding to the derived companion mass ($\sim0.7 \Msun$). Since the typical photometric uncertainty on the single measurements is $\sim0.08$ mags, such a small magnitude modulation cannot be detected with our dataset. Magnitude modulations comparable to or larger than our typical photometric uncertainty are expected only for $i\geq30^{\circ}$, corresponding to companion masses $\leq0.2 \Msun$ (see an example in Figure~\ref{curva}), which have been excluded by our analysis. This seems to further support the conclusion that the system is seen at a small inclination angle. However,  this is a very basic model and the addition of processes like irradiation from the primary star, truncated disk,  etc.. can modify the light curve shape. Hence, deeper observations are needed before drawing solid conclusions about the optical variability of the system.}

The spin and orbital properties of \sax are quite similar to those generally observed in redback pulsars. This, combined with the periodical occurrence of outbursts, might suggest that this system is a tMSP whose radio pulsed emission has not been revealed yet (see Section~\ref{intro}). However, here we presented some observational evidence suggesting {\ an on-going mass transfer during the quiescence state. This is not expected in the radio pulsar state of tMSPs \citep[see e.g.][]{archibald09,pallanca13} and suggests that in the case of \sax the radio emission mechanism is not active and thus that the system is not a tMSP. The $H\alpha$ emission detected in \sax clearly indicates that this system behaves as a typical low-mass X-ray binaries in quiescence, with mass transfer currently on-going and a possible residual accretion disk still present around the neutron star. This shows that not all accreting neutron stars with main sequence companions and orbital parameters similar to redbacks behave as tMSPs.}

\newpage

\section{CONCLUSIONS}
We presented the optical identification and characterization of the AMXP \sax during quiescence. We identified a possibile counterpart (\com) in a star located at only $\sim0.15\arcsec$ from the X-ray nominal position. This star, although being located along the cluster main sequence, shows an excess in the F656N filter, thus implying the presence of $H\alpha$ emission.  {\  We discussed the physical properties of the companion star and show that it has a mass of $0.70 \Msun - 0.83M_{\odot}$, an effective temperature of $5250 K$ and it is filling, or even overflowing, its Roche-Lobe radius of $0.88\pm0.02 R_{\odot}$. This mass, combined with the binary system mass function and assuming canonical neutron star masses, implies that the binary system is observed at a very low inclination angle ($\sim8^{\circ}-14^{\circ}$)}. This can also explain the absence of a significant magnitude variability. The EW of the $H\alpha$ emission has been evaluated to be of about 20 \AA, which corresponds to a mean mass transfer rate during quiescence of $\sim10^{-10} \, \, M_{\odot} \, yr^{-1}$. {\ The possibility of on-going mass transfer and residual accretion disk around the neutron star during quiescence states  implies that the radio pulsar is not reactivated yet. Hence \sax is probably not a tMSP and its behavior during quiescence is comparable with that commonly observed in classical quiescent low-mass X-ray binaries, even tough its orbital and spin parameters are very similar to those observed for redback millisecond pulsars. This directly implies that not all the redback-like AMXP with main sequence companions are tMSPs. For some reasons, \sax pulsar is not able to turn-on the radio emission during the quiescence state, at odds with what happens for tMSPs. The reasons behind this are still obscure. Intriguingly, it is worth noticing that \com has a mass larger than that measured for the companion stars of redbacks and tMSPs \citep[$0.2\Msun-0.4 \Msun$, see, e.g.,][]{breton13,mucciarelli13,bellm16}. Therefore the companion mass could be one of the ingredients to understand why this redback-like AMXP is not behaving like a tMSP.} 



\section{Acknowledgement}

{ We thank the anonymous referee for the prompt reading of the manuscript and for the useful comments.}
A.P. acknowledges support from a Netherlands Organization for Scientific Research (NWO) Vidi Fellowship.

\clearpage


\end{document}